\documentclass[runningheads]{svmult}
\usepackage{makeidx}   
\usepackage{graphicx}  
\usepackage{subeqnar}  
\usepackage{multicol}  
\usepackage{physprbb}  
\def\Journal#1#2#3#4{{#1} {\textbf{#2}}, #3 (#4)}

\def\NPB{Nucl. Phys. \textbf{B}}
\def\PLB{Phys. Lett. \textbf{B}}
\def\PRL{Phys. Rev. Lett.}
\def\PRD{Phys. Rev. \textbf{D}}

\begin{document}
\title*{The Rydberg-Atom-Cavity Axion Search}
\toctitle{The Rydberg-Atom-Cavity Axion Search}
%
%
\titlerunning{The Rydberg-Atom-Cavity Axion Search}
%
\author{K. Yamamoto\inst{1}
\and M. Tada\inst{2}
\and Y. Kishimoto\inst{2}
\and M. Shibata\inst{2}
\and K. Kominato\inst{2}
\and T. Ooishi\inst{2}
\and S. Yamada\inst{3}
\and T. Saida\inst{2}
\and H. Funahashi\inst{3}
\and A. Masaike\inst{4}
\and S. Matsuki\inst{2}}
\authorrunning{K. Yamamoto et al.}
%
%
\institute{Department of Nuclear Engineering,
Kyoto University, Kyoto 606-8501, Japan
\and Nuclear Science Division,
Institute for Chemical Research, Kyoto University, \\
Uji, Kyoto 611-0011, Japan
\and Department of Physics,
Kyoto University, Kyoto 606-8501, Japan
\and Faculty of Engineering, Fukui University of Technology,
Fukui 910-8505, Japan}

\maketitle              

\renewcommand{\thefootnote}{{ }}
\footnote{Invited talk presented at the Dark2000, Heidelberg, Germany,
10-15 July, 2000}

\begin{abstract}
We report on the present progress
in development of the dark matter axion search experiment
with Rydberg-atom-cavity detectors in Kyoto, CARRACK I and CARRACK II.
The axion search has been performed with CARRACK I in the 8 \% mass range
around $ 10 \mu {\rm eV} $, and CARRACK II is now ready
for the search in the wide range $ 2 \mu {\rm eV} - 50 \mu {\rm eV} $.
We have also developed quantum theoretical calculations
on the axion-photon-atom system in the resonant cavity
in order to estimate precisely the detection sensitivity
for the axion signal.
Some essential features on the axion-photon-atom
interaction are clarified, which provide the optimum experimental
setup for the axion search.
\end{abstract}

\section{Introduction}
The axion \cite{axion} in the mass range
$ m_a = 1 \mu {\rm eV} - 1 {\rm meV} $
is one of the most promising candidates
for the non-baryonic dark matter in the universe
\cite{dm}.
The search for dark matter axions is, however,
a quite difficult task due to their extremely weak interactions
with ordinary matter.
The basic idea for dark matter axion search is to convert
axions into microwave photons in a resonant cavity
under a strong magnetic field via Primakoff process,
as originally proposed by Sikivie
\cite{axion-conversion}.
Pioneering experiments were made before
with amplification-heterodyne-method
\cite{old-exp}.
Recently, some results of an advanced experiment
by the US group have been reported,
excluding the KSVZ axion with mass $ 2.9 \mu {\rm eV} - 3.3 \mu {\rm eV} $
as the dark matter in the halo of our galaxy
\cite{US-exp}.

We have proposed a quite efficient scheme for dark matter axion search,
where Rydberg atoms are utilized to detect the axion-converted photons
\cite{acb-qm}.
Then, based on this scheme we have developed
ultra-sensitive Rydberg-atom-cavity detectors,
CARRACK I and II (Cosmic Axion Research with Rydberg Atoms
in resonant Cavities in Kyoto)
\cite{acb-exp}.
We here report on the present progress
in development of the dark matter axion search experiment
with Rydberg-atom-cavity detectors.
The axion search in the mass range $ 2350 {\rm MHz} - 2550 {\rm MHz} $,
about 8 \% around $ 10 \mu {\rm eV} $, has been performed
with the prototype detector CARRACK I.
Then, based on the performance of CARRACK I,
the new large-scale apparatus CARRACK II is now ready
for the axion search in the wide mass range
$ 2 \mu {\rm eV} - 50 \mu {\rm eV} $.

\section{Rydberg-atom-cavity detector}
The principle of the present experimental method
is schematically shown as follows.
\vspace{-0.5cm}
\[
\begin{array}{l}
\hspace{0.7cm}
\left[
\begin{array}{c} {\bf conversion} \\ {\bf cavity} \end{array}
\right]
\hspace{0.3cm}
\left[
\begin{array}{c} {\bf detection} \\ {\bf cavity} \end{array}
\right]
\\
{ }
\\
{\bf \langle \ axions \ \rangle}
\begin{array}{c} B_0 \\ \Longleftrightarrow
\\ g_{a \gamma \gamma} \end{array}
{\bf \langle \ photons \ \rangle}
\begin{array}{c} d \\ \Longleftrightarrow \\ \Omega_N \end{array}
{\bf \langle \ atoms \ \rangle}
\end{array}
\begin{array}{l}
{ }
\\
{ }
\\
{ }
\\
{ }
\\
{ }
\\
- \! \! \! - | e \rangle \rightarrow {\mbox{\it selective field ionization}}
\\ \hspace{.15cm}
\updownarrow \\ - \! \! \! - | g \rangle
\\ \hspace{.15cm}
\uparrow {\mbox{\it laser excitation}}
\\ \hspace{.15cm}
{\mbox{ground state (Rb)}}
\end{array}
\]
The dark matter axions are converted into photons
under a strong magnetic field $ B_0 $ in the conversion cavity
through the axion-photon-photon coupling $ g_{a \gamma \gamma} $.
These axion-converted photons are transferred to the detection cavity,
and interact with Rydberg atoms passed through the cavity
due to the electric dipole transition $ d $
providing the collective atom-photon coupling $ \Omega_N $.
The Rydberg atoms are initially prepared to the lower state $ | g \rangle $.
Then, the atoms excited to the upper state $ | e \rangle $
by absorbing the axion-converted photons are detected quite efficiently
with the selective field ionization method
\cite{Gallagher,sfi}
after exiting the cavity.
The background noise in this method is predominantly brought
by the thermal photons in the cavity
which can also excite the Rydberg atoms.
It can be reduced substantially by cooling the whole apparatus
down to about $ 10 {\rm mK} $,
attaining a significant signal-to-noise ratio.
Therefore, the Rydberg-atom-cavity detector,
which is free from the amplifier noise by itself,
is expected to be quite sensitive for the dark matter axion search.

\begin{figure}[h]
\begin{center}
\includegraphics[width=.7\textwidth]{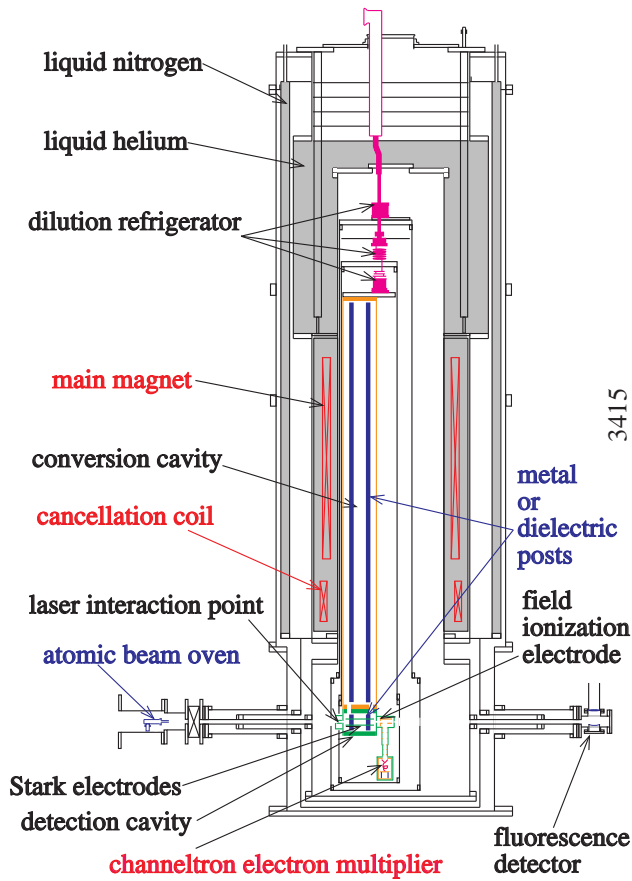}
\end{center}
\caption{The layout of CARRACK II.}
\label{fig:CARRACKII}
\end{figure}

The layout of the actual apparatus, CARRACK II, is shown
in Fig. \ref{fig:CARRACKII}.
It mainly consists of 6 parts;
superconducting magnets, coupled cavities (conversion and detection),
dilution refrigerator, atomic beam system, laser excitation system
and selective field ionization (sfi) system.

The magnet system consists of 2 superconducting solenoid coils.
One is the main coil which can produce the magnetic flux density
of $ 7 \ {\rm T} $ at the center.
The other is the cancellation coil
which is set at the lower side of the main coil
to reduce the magnetic field in the region of the detection cavity
to less than $ 900 \ {\rm Gauss} $
(the critical magnetic flux density of superconducting niobium
is $ 1200 \ {\rm Gauss} $).

The conversion cavity is made of oxygen-free high-conductivity copper.
The detection cavity and the sfi housing are made of niobium
to prevent the penetration of the magnetic field into the inside
of these area.  This is to avoid complicated level splitting and shift
due to Zeeman effect to the atoms.
The cavity mode is the cylindrical $ {\rm TM}_{010} $ mode
and the cavity resonant frequency is tuned over 25 \%
by the metal and aluminum-oxide posts
\cite{posts}.
The transition frequency of the Rydberg atom should approximately
coincide with the resonant frequency of the cavity.
It is tuned roughly by choosing
a state with appropriate principal quantum number $ n $,
and finely by applying an electric field with Stark electrodes
in the detection cavity.
The atoms in the $ s_{1/2} $ and $ p_{1/2} $ states are Stark shifted
as $ - \alpha_0 E^2 / 2 $ with a certain constant $ \alpha_0 $
and the applied electric field $ E $.
We have measured the scaler polarizability $ \alpha _0 $
of the relevant levels for a wide range of $ n $ (60 to 150)
and hence obtained detailed systematic information on this tuning.

The dilution refrigerator (Oxford Kelvinox 300)
is used to cool the cavity system down to the low enough temperature
$ T_c \sim 10 {\rm mK} $.

We use two kind of atomic beam source systems.
One consists of an ion source and a charge exchange cell;
atoms are ionized once and after accelerated, they are neutralized
in the charge exchange cell to get higher-velocity atomic beam
with kinetic energy $ 10 {\rm eV} - 100 {\rm eV} $.
In the other system, a thermal atomic oven is used
to get lower-velocity atomic beam
with $ 350 {\rm m}{\rm s}^{-1} - 450 {\rm m}{\rm s}^{-1} $.
As discussed later, the atomic velocity should be changed
roughly proportional to the axion mass
in order to attain the optimal sensitivity.
Hence, with these two systems a wide range of the atomic beam velocity
can be covered to meet the need from the experimental situation.

The Rydberg states are produced with 2-step laser system.
At the first step, atoms are excited from the ground state
to $ 5 p_{3/2} $ state by a diode laser with wavelength
$ 780.24 \ {\rm nm} $, and then to $ n s_{1/2} $ state
($ n \sim 110 $) by a ring dye laser.
The wavelength of the ring dye laser is varied from $ 479.13 \ {\rm nm} $
to $ 479.65 \ {\rm nm} $ depending on the axion mass.
The ring dye laser is pumped by a Kripton ion laser.
The transition from the $ n s_{1/2} $ state
to the $ n p_{1/2} $ or $ n p_{3/2} $ state
is used to absorb the axion-converted photons.

The atoms excited to the upper state by absorbing microwave photons
are selectively ionized by applying a pulsed electric field.
By taking an appropriate slew rate of the pulsed electric field,
the difference of the ionization field values
between the upper and lower atomic states becomes large enough.
This enables us to ionize selectively the upper state
with quite good efficiency
\cite{sfi}.

\section{Sensitivity for the dark matter axions}

The Rydberg-atom-cavity detector
is an ultra-sensitive single-photon counter.
In the theoretical point of view, it is treated
as a quantum system of interacting oscillators
with dissipation which represent appropriately
the axions, photons and atoms in the cavity.
We have developed quantum theoretical calculations
on the axion-photon-atom system in the resonant cavity
in order to estimate precisely the detection sensitivity
for dark matter axions
\cite{acb-qm}.
These calculations are made by taking into account
appropriately the actual experimental situations
such as the motion and uniform distribution of Rydberg atoms
in the incident beam as well as
the spatial variation of the electric field in the cavity.
We here recapitulate the essential results,
and show how the relevant experimental parameters
should be adjusted to attain the optimal sensitivity.

The characteristic properties of axions, photons and atoms
are listed as
\[
\begin{array}{l}
{\bf axions}:
\\ \hspace{0.5cm}
m_a \sim 10^{-5} {\rm eV} = 2.4 {\rm GHz} , \
\rho_a \sim \rho_{\rm halo} = 0.3 {\rm GeV}/{\rm cm}^3 , \
\beta_a \sim 10^{-3} ,
\\ \hspace{0.5cm}
\lambda_a \simeq ( 2 \pi \hbar / \beta_a m_a ) \sim 100 {\rm m} , \
{\bar n}_a \simeq \lambda_a^3 ( \rho_a / m_a ) \sim 10^{26} ,
\\ \hspace{0.5cm}
\gamma_a \sim \beta_a^2 m_a / \hbar \sim 10^{-11} {\rm eV} / \hbar ,
\\
{\bf photons}:
\\ \hspace{0.5cm}
{\bar n}_c
= \left( {\rm e}^{\hbar \omega_c / k_{\rm B} T_c} - 1 \right)^{-1}
\sim 10^{-5} , \ T_c \sim 10 {\rm mK} ,
\\ \hspace{0.5cm}
\gamma_c \equiv \gamma = \omega_c / Q \sim 10^{-10} {\rm eV} / \hbar ,
\\
{\bf atoms}:
\\ \hspace{0.5cm}
{\bar n}_b = 0 \ {\mbox{(initially in the lower state)}} , \
\gamma_b \sim 10^{-13} {\rm eV} / \hbar \
( \tau_b \sim 10^{-3} {\rm s} ) .
\end{array}
\]
The relevant experimental parameters are also given as
\[
\begin{array}{l}
\hspace{0.5cm}
{\mbox{\it magnetic flux}}: B_0 \sim 7 {\rm T} , \
{\mbox{\it quality factor}}: Q \sim 3 \times 10^4 ,
\\ \hspace{0.5cm}
{\mbox{\it single atom-photon coupling}}:
\Omega \sim 5 \times 10^3 {\rm s}^{-1} ,
\\ \hspace{0.5cm}
{\mbox{\it atomic velocity}}: v \sim 350 {\rm m}{\rm s}^{-1} , \
{\mbox{\it atomic passing distance}}: L \sim 0.2 {\rm m} ,
\\ \hspace{0.5cm}
{\mbox{\it atomic beam intensity}}:
I_{\rm Ryd} = N (v/L) \sim 10^5 {\rm s}^{-1} ,
\\ \hspace{0.5cm}
{\mbox{\it number of atoms}}: N \sim 10^2 ,
\\ \hspace{0.5cm}
{\mbox{\it scanning frequency step}}:
\Delta \omega_c \sim \beta_a^2 m_a / \hbar \sim 5 {\rm kHz} .
\end{array}
\]
These parameter values are optimum for the dark matter axion search
as shown below.

In order to estimate the sensitivity for the dark matter axions,
we need to calculate the counting rates of the excited atoms per unit time
at the exit of cavity which are due to the axion-converted photons
and the thermal background photons, respectively.
(See Ref. \cite{acb-qm} for the details.)
The signal and noise rates are calculated
in terms of the atomic velocity $ v $
and the densities of the atoms in the upper state at the exit of cavity
$ {\bar \rho}_b^{[a]} (L) $ and $ {\bar \rho}_b^{[ \gamma ]} (L) $:
\begin{equation}
R_s = v {\bar \rho}_b^{[a]} (L) , \
R_n = v {\bar \rho}_b^{[ \gamma ]} (L) .
\end{equation}
The resonant absorption of the microwave photons by the Rydberg atoms
is determined by the atomic damping rate
as well as the atomic transition frequency fine-tuned with Stark effect.
It should be noted that the effective atomic damping rate
may be larger than the original one $ \gamma_b $
due to the collective atom-photon coupling
$ \Omega_N $ and the finite atomic transit time $ t_{\rm tr}$,
which are given by
\begin{eqnarray}
\Omega_N &=& {\sqrt N} \Omega ,
\\
t_{\rm tr} &=& L/v .
\end{eqnarray}
The effective atomic width is roughly estimated as
\begin{equation}
{\bar \gamma}_b \sim \gamma_b + ( \Omega_N / \gamma )^2 \gamma + v/L .
\end{equation}
In the weak region of atomic beam intensity $ I_{\rm Ryd} $
providing small enough $ \Omega_N $,
the signal and noise rates increase
with $ I_{\rm Ryd} \propto N \propto \Omega_N^2 $.
Then, for certain beam intensity
$ I_{\rm Ryd} = {\bar I}_{\rm Ryd} $, where the condition
\begin{equation}
{\bar \gamma}_b \sim \gamma_a \sim \beta_a^2 m_a / \hbar
\end{equation}
is satisfied with
\begin{equation}
{\bar \Omega}_N \sim ( \gamma_a \gamma )^{1/2} , \ 
v \sim L \gamma_a ,
\end{equation}
the signal rate is maximized to be $ {\bar R}_s $,
and the noise rate almost reaches the asymptotic value $ {\bar R}_n $.
These counting rates in the optimum case
are roughly estimated as
\begin{eqnarray}
{\bar R}_s & \sim &
(v/L) ( \gamma / \gamma_a ) ( \kappa / \gamma )^2 {\bar n}_a ,
\\
{\bar R}_n & \sim & (v/L) {\bar n}_c ,
\end{eqnarray}
which are proportional to the number of axions
and the number of thermal photons, respectively.
The effective axion-photon coupling in the resonant cavity
under the strong magnetic field is calculated
from the original $ g_{a \gamma \gamma} $ coupling
\cite{acb-qm} as
\begin{eqnarray}
\kappa & = & 4 \times 10^{-26} {\rm eV} \hbar^{-1}
\left( g_{a \gamma \gamma} / 1.4 \times 10^{-15} {\rm GeV}^{-1} \right)
\left( G B_0 / 4 {\rm T} \right)
\nonumber \\
& \times &
\left( \beta_a m_a / 10^{-3} \times 10^{-5} {\rm eV} \right)^{3/2}
\left( V_1 / 5000 {\rm cm}^3 \right)^{1/2} ,
\end{eqnarray}
where $ V_1 $ and $ G $ are the volume and form factor
of the cavity system, respectively.

\begin{figure}[h]
\begin{center}
\includegraphics[width=.7\textwidth]{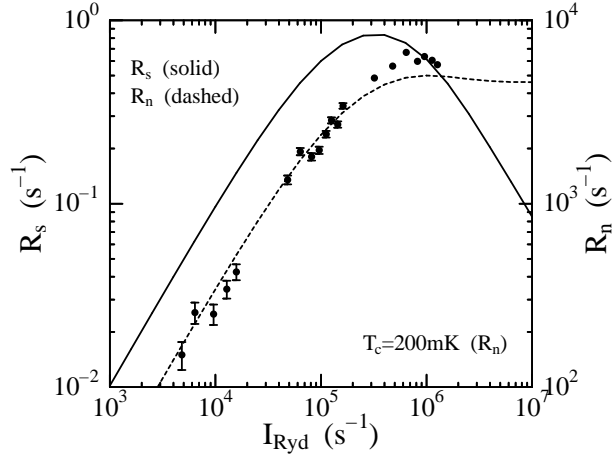} \vspace{-0.5cm}
\end{center}
\caption{The theoretical estimates of the signal and noise rates are shown.
The experimental data of the thermal photon noise
are also presented for comparison.}
\label{fig:th-exp}
\end{figure}

In Fig. \ref{fig:th-exp}, we show the theoretical estimates
of the signal rate $ R_s $ (solid) and noise rate $ R_n $ (dashed).
The signal and noise rates really exhibit the characteristic behaviors
with respect to the atomic beam intensity as mentioned above.
The preliminary experimental data of the thermal photon noise
(dots with error bars) are also presented for comparison.
The theoretical estimates are indeed in good agreement
with the experimental data being proportional
to the number of thermal photons $ {\bar n}_c ( \omega_c / T_c ) $.
This indicates that the thermal photon noise provides
an efficiency calibration for the Rydberg-atom-cavity detector.
Specifically, the optimum beam intensity for the dark matter axion search
can be determined rather accurately by detecting the thermal photon noise.
The signal rate is in fact maximized where the noise rate
turns to be saturated.

The sensitivity for the axion search at $ m \sigma $ level
is estimated with the signal and noise rates
$ R_s \sim {\bar R}_s $ and $ R_n \sim {\bar R}_n $.
The one-step measurement time and the total scanning time
over a 10 \% frequency range are calculated respectively by
\begin{eqnarray}
\Delta t &=& m^2 ( 1 + R_n / R_s ) / R_s ,
\\
t_{\rm tot} &=& ( 0.1 \omega_c / \Delta \omega_c ) \Delta t .
\end{eqnarray}
We have estimates at $ 3 \sigma $ level for the DFSZ axion
with mass around $ 10 \mu {\rm eV} $,
\[
\Delta t \sim 100 {\rm s} , \ t_{\rm tot} \sim 100 {\rm days} ,
\]
by taking the relevant experimental parameter values as listed so far.
The sensitivity is of course much better for the KSVZ axion.

\section{Status and prospect}

We have made so far extensive research and development
in the dark matter axion search with Rydberg-atom-cavity detector.

In the theoretical part, we have developed
the quantum theoretical formulations and calculations
for the axion-photon-atom interaction in the resonant cavity.
They provide the precise estimate of the sensitivity
for the dark matter axions, specifying the optimum setup
for the relevant experimental parameters
such as the atomic beam velocity and intensity.
Then, by using these calculations
we can determine the bound on the axion-photon-photon coupling
$ g_{a \gamma \gamma} $ from the experimental data.

Experimentally, we have searched for the dark matter axions
in the mass range $ 2350 {\rm MHz} - 2550 {\rm MHz} $
around $ 10 \mu {\rm eV} $ with the prototype detector CARRACK I.
The experimental parameters are taken as
$ T_c = 12 {\rm mK} - 15 {\rm mK} $, $ \Delta \omega_c = 10 {\rm kHz} $,
$ Q = 4 \times 10^4 $, $ \Delta t = 300 {\rm s} $,
$ I_{\rm Ryd} = 5 \times 10^5 {\rm s}^{-1} $,
$ v = 350 {\rm m}{\rm s}^{-1} $.
The theoretical calculations indicate that
the sensitivity with these parameter values
exceeds the limit of KSVZ axion,
$ g_{a \gamma \gamma}^2 < 1.4 \times 10^{-29} {\rm GeV}^{-2} $.
The actual limit will be placed soon
after making some more detailed calculations and checks.

We have also made various developments for the detection apparatus;
$ \bullet $ significant improvement in the selective field ionization
with pulsed electric field,
$ \bullet $ keeping the good performance of apparatus
in a long-term run at the very low temperature $ \sim 10 {\rm mK} $,
$ \bullet $ sufficient cancellation of magnetic field
in the detection cavity made of niobium,
$ \bullet $ precise tuning of the resonant frequencies
of the coupled cavities and the atomic transition frequency
with Stark shift,
$ \bullet $ improvement of the atomic beam source
providing better quality beam, and so on.

Now, we are ready for the search in the wide range
$ 2 \mu {\rm eV} - 50 \mu {\rm eV} $
with the large-scale apparatus CARRACK II.
We will reach in a few years the DFSZ limit
throughout this axion mass range.

\section*{Acknowledgments}
This research was partly supported
by a Grant-in-Aid for Specially Promoted Research
by the Ministry of Education, Science, Sports and Culture,
Japan under the program No. 09102010.

%

\end{document}